\documentclass[prl,twocolumn,floatfix]{revtex4}
\newcommand{\be}{\begin{eqnarray}}
\newcommand{\ee}{\end{eqnarray}}
\newcommand{\ket}[1]{\ensuremath{\left| {#1} \right>}}
\newcommand{\bra}[1]{\ensuremath{\left< {#1} \right|}}

\newcommand{\nbar}{\bar{n}}
\newcommand{\upup}{\uparrow\uparrow}
\newcommand{\updown}{\uparrow\downarrow}
\newcommand{\downup}{\downarrow\uparrow}
\newcommand{\downdown}{\downarrow\downarrow}

\usepackage{graphicx}
\usepackage{amsmath}
\usepackage{amssymb}

\begin{document}

\title{Deterministic entanglement and tomography of ion spin qubits}

\author{J.~P.~Home, M.~J.~McDonnell, D.~M.~Lucas, G.~Imreh, B.~C.~Keitch, D.~J.~Szwer, N.~R.~Thomas, S.~C.~Webster, D.~N.~Stacey and A.~M.~Steane.}

\affiliation{Centre for Quantum Computation, Clarendon Laboratory, Department of Physics,
University of Oxford, Parks Road, Oxford OX1 3PU, UK}

\begin{abstract}
We have implemented a universal quantum logic gate between qubits
stored in the spin state of a pair of trapped $^{40}$Ca ions. An
initial product state was driven to a maximally entangled state
deterministically, with 83\% fidelity. We present a general
approach to quantum state tomography which achieves good
robustness to experimental noise and drift, and use it to measure
the spin state of the ions. We find the entanglement of formation is
$0.54$.
\end{abstract}

\maketitle

In recent years, entanglement has been detected and used in a
variety of physical systems, but the number in which it can be
created under deterministic control with good fidelity remains
small. We present the implementation of a universal 2-qubit
quantum logic gate in a trapped ion experiment, demonstrating
deterministic entanglement with 83\% fidelity.  Tomography was
performed using a method robust against noise sources of practical
relevance in the experiments. Our results illustrate the relative
insensitivity of the gate to the initial temperature of the ions.

The logic gate exploits the oscillating force
method \cite{03Leibfried, 00Milburn2, 00Sorensen1, 01Wang, 99Sorensen}.
Our work differs from previous trapped ion
experiments in the physical nature of the qubit, the trapping
parameters, the light fields used to implement logic at the ions,
and the laser sources, as follows.

First, each qubit is stored in a pure spin-half system, the
spin of a $^{40}$Ca ion its ground state. Previous work used
hyperfine
structure \cite{98Turchette,03Leibfried,05Haljan,05Leibfried} or a pair
of electron orbitals \cite{03SchmidtKaler,05Haffner, footnote1}. The spin-half system is
convenient in its simplicity. The absence of `spectator levels',
i.e. other states nearby in energy, reduces the movement
of population out of the controlled Hilbert space. This
process, sometimes called leakage error, is not
directly correctable by quantum error correction \cite{BkLo,BkNielsen}.
Spectator levels also exacerbate decoherence by photon
scattering during the gate operation.

Next, our trap is comparatively large \cite{04Lucas}, so electric
field noise from fluctuating patch potentials and Johnson noise in
the electrodes is very small \cite{00Turchette,04Deslauriers}. We have measured
very low heating rates, of order a few phonons per second, in the
trap.

We use a single laser field to apply all the operations (single
ion rotations and the 2-qubit gate), changing only the amplitude
and frequency to switch from one operation to another.
This reduces the experimental complexity, and opens
the possibility of future work in which the different parts of a
pulse sequence could be brought together in a single chirped and
shaped pulse.

Finally, the laser sources in our apparatus are all small
semiconductor diodes, which do not require frequency doubling and which could in principle be packaged on a
small optical chip. We thus begin to address the optical part of
the challenge to scale this type of system up to a powerful
quantum computer\cite{98Wineland2,02Kielpinski,04Steane2}.

Let the spin-up/down states $\ket{\uparrow}, \ket{\downarrow}$ be
the computational basis states of a qubit. We implement the
2-qubit controlled-phase gate
$Z_1(\phi_1) Z_2(\phi_1) G(\Psi)$ where $Z_k(\phi_1)$ is a rotation
of qubit $k=1,2$ by $\phi_1$ about the quantization axis, and $G(\Psi)$ is
a two-qubit operator represented in the computational basis
$\ket{\uparrow\uparrow},\ket{\uparrow\downarrow},\ket{\downarrow\uparrow}
,\ket{\downarrow\downarrow}$ by the diagonal matrix ${\rm
diag}(1,\exp(i \Psi),\exp(i \Psi),1)$. For $\Psi = \pi/2$, or an
odd multiple of $\pi/2$, this is equivalent to controlled-not up
to single qubit rotations.

The gate mechanism is described in \cite{03Leibfried}.
A spin-dependent force oscillating at frequency $\omega$ is
applied to the pair of ions. Let $f_k(m)$ be the complex amplitude
of the force on ion $k$ when it is in spin state
$m=\uparrow,\downarrow$. The COM (stretch) mode is
excited by the sum (difference) force $f_1(m_1) \pm f_2(m_2)$
respectively. The effect on the quantum harmonic motion
is simply to displace the state in its $z$--$p$ phase
space (as long as the position-dependence of the force is
negligible during the displacement \cite{ourcat}). By choosing $\omega$ close to
one of the normal mode frequencies one can enhance the
displacement and also simplify the dynamics, because then the
excited mode simply describes a circle in phase space, which
closes after time $2\pi/\delta$, and the excitation of the
other mode can be neglected. This motion causes the system to
acquire a phase $\phi_L$ proportional to the area of the loop in
phase space.

Many arrangements are possible. We adopt $f_1(m) = f_2(m)$, then the stretch
mode is excited only for anti-aligned spin states. By tuning close
to the stretch mode frequency $\omega_s$, i.e. $|\delta| \ll \omega_s -
\omega_c$ where $\delta \equiv \omega - \omega_s$ and $\omega_c$ is the COM mode frequency,
we then have $\Psi = \phi_L = (\pi/2)(\Omega_f/\delta)^2$
where $\Omega_f = |f(\uparrow)-f(\downarrow)| z_{0s}/\hbar$,
$z_{0s} = (\hbar /(4 M \omega_s))^{1/2}$ and
$M$ is the mass of one ion \cite{footnote2}.

To measure the spin state of the ion pairs, we developed the
following quantum state tomography method. Tomography in general
requires the accumulation of a sufficient set $\{ M_i \}$ of
measurements of a given state $\rho$ (many copies of which
are needed), in order to allow $\rho$ to be
reconstructed to some desired accuracy from $\{ M_i \}$.
Usually rotations are applied to the system which is then
measured in a fixed basis.

This problem has been studied in various settings
and, for example,
sets of $\{ M_i \}$ of minimal size, or designed for
certain experimental situations, have been discovered \cite{BkParis}.
Our method is designed to be robust
against noise and drift problems, and to be
convenient in a fairly broad range of experimental settings.
We achieve this by a judicious choice of rotation
angles and by accumulating data sets in a form that should be
well-fitted by sums of orthogonal sinusoidal functions of known period.
The least squares fit is robust because the functions
in the fit are orthogonal, it averages the data
over many timescales, and many noise processes affect the
residuals not the fitted parameter values.

Let $R(\theta,\phi)$ be a rotation of a single qubit through
$\theta$ on the Bloch sphere about an axis of azimuthal angle
$\phi$. To analyse a given density matrix $\rho$,
the qubits are first rotated and then measured in the $\ket{\uparrow},\ket{\downarrow}$
basis, then $\rho$ is re-prepared and one repeats with either the same or new rotation
angles, until a sufficient set $\{ M_i \}$ has been acquired.
Experimentally $\theta$ depends on a pulse area, $\phi$ on the
phase of an r.f. oscillator. The latter is more easily adjusted than
the former to a variety of precisely known values. Therefore we
only use two values of $\theta$, but many values of $\phi$, see
fig. 2. For clarity we give details for the case of
two qubits both undergoing the same rotation, but the principles
apply more generally. In this restricted case a general $\rho$ can
be almost, but not fully reconstructed.

Write $\rho =
\sum_{i,j=0}^3 c_{ij} \hat{\sigma}_i \otimes \hat{\sigma}_j$ where $\hat{\sigma}_i$
are the Pauli spin operators (starting with $\hat{\sigma}_0=I$ the identity).
Let $P(m_1 m_2)$ be the populations of $\rho$ and $P_R(m_1,m_2)$ be
the populations after the rotation $R(\theta,\phi)$ has been applied to
both qubits. Then
\be
P_R(\upup) &=&
  a_{\uparrow\uparrow}
+ b_{\uparrow\uparrow} \cos(\phi)
+ c_{\uparrow\uparrow} \sin(\phi) \nonumber \\
& & + d_{\uparrow\uparrow} \cos(2\phi)
+ e_{\uparrow\uparrow} \sin(2\phi)  \label{fit}
\ee
and similarly for $\updown,\downarrow\uparrow$,
where the coefficients $a$--$e$
are simple sinusoidal functions of $\theta$.
By choosing different values of $\theta$ and
$\phi$, 12 independent real numbers can be extracted from the
population measurement outcomes. Therefore 3 of the 15 independent
real numbers which fully characterize $\rho$ are not available
when both qubits undergo the same rotation.

The tomography method consists of accumulating data at given
$\theta$ as a function of $\phi$, then fitting eq.
(\ref{fit}) and its partners for $P_R(\updown), P_R(\downup)$ to the
data, with the $a$--$e$ coefficients as fitted parameters, c.f. fig. 2.
After using two $\theta$ values, the density matrix
coefficients $c_{ij}$ can be found from $\theta$ and $a$--$e$
(with some redundancy).
The inferred matrix $\rho^M$ is not guaranteed to be
positive definite. One can handle this by any suitable approach,
for example by searching among physical density matrices $\rho^P$
for the one which most closely matches $\rho^M$ by some measure.
We adopt this ``maximum likelihood'' method using the cost
function $\sum_{i,j} | \rho^P - \rho^M |^2_{ij}$ where $\rho^P$ is
positive definite \cite{01James}.

The apparatus consists of a linear r.f. Paul trap and laser system
largely as described in \cite{04Lucas}. We load single or pairs
of $^{40}$Ca ions by photoionisation. A
6.3 MHz r.f. field provides radial confinement, with radial COM
frequency $\simeq 1\,$MHz. Axial confinement
is provided by d.c. endcaps, tuned as described below.
397 nm and 866 nm lasers are used for Doppler cooling, fluorescence
detection and spin state preparation (with 99\% fidelity) by optical pumping.
A further pair of lasers performs spin-state-selective shelving
to $D_{5/2}$, which allows the fluorescence to be used to read out the spin
state (with $\sim 90$\% fidelity), see \cite{04McDonnell1}.
The ions are cooled in 3 dimensions by Doppler cooling, and along $z$
by Raman sideband cooling to $\nbar_{\rm com,str} \simeq 0.2, 0.2$,
measured as in ref. \cite{98King}.

The light field used for pulsed sideband cooling and all the coherent
operations is a walking wave created by laser beams $A,B$ propagating at
$\theta_L = 58.9^{\circ}$ to each other, with difference wavevector
$\Delta k = 4\pi \sin (\theta_L/2) / \lambda$ directed along the trap axis $z$.
A quantization axis is
set by a weak magnetic field directed at $3^{\circ}$ to beam $B$
(and $\theta_A = 62^{\circ}$ to beam $A$).
All these axes are horizontal. To reduce sensitivity to
pulse area we balance the a.c. Stark shifts on
$\ket{\uparrow}$, $\ket{\downarrow}$ from either beam acting alone,
using Ramsey interferometry. This results in polarization very close to
linear \cite{footnote3},
which we make at angle $\beta$ to the vertical for $A$, and horizontal
for $B$. Each beam illuminates both ions equally.
They are derived from the same laser, at a frequency detuned
$\Delta_L /2\pi = 30$ GHz above the $S_{1/2}-P_{1/2}$ transition. A precise frequency
difference $\omega$ is introduced between them by acousto-optic modulation.

The Zeeman splitting of $\ket{\uparrow}$ from $\ket{\downarrow}$
is $\omega_0 = 2\pi \times 4800\,$kHz. At $\omega = \omega_0$
spin-flip transitions are resonantly driven by the $\pi$ component of $A$ combined
with the $\sigma^-$ component of $B$. Let $\Omega_c$ be
the Rabi frequency for this `carrier' Raman process. The oscillating force is created by
the periodic light shift from the $\sigma^{\pm}$ components of the
walking wave.
The coupling strength $\Omega_s$ is given by
$\Omega_s = \sqrt{2}\Omega_c \epsilon_+/\epsilon_\pi$, where $\epsilon_+, \epsilon_\pi$ are
polarization amplitudes. For our geometry, $\sqrt{2} \epsilon_+/\epsilon_\pi =
\cot{\theta_A}/\cos(\Delta \phi/2)$ and
$\Omega_f = 2 \eta \Omega_s \sin(\Delta\phi/2)$,
where $\eta = \Delta k z_{0s}$ is the stretch mode Lamb-Dicke parameter, and
$\Delta\phi$ is the phase angle between the forces, given by
$\tan(\Delta\phi/2) = (\cos \theta_A \tan \beta)^{-1}$.

During the force pulse at $\omega \simeq \omega_s$,
the carrier process is driven off-resonantly at detuning
$\omega_0 \pm \omega$. For $\omega_0 \pm \omega \gg
\Omega_c$, the qubit states are pushed apart by a light
shift
\be
\Delta_c = \frac{\Omega_c^2}{2} \left[ \frac{1}{\omega_0 + \omega} + \frac{1}{\omega_0 - \omega} \right].
\ee
This is of order 4~kHz in our experiments, and it gives rise to the single-qubit
rotations $Z_k(\phi_1)$ in the gate, with $\phi_1 = \Delta_c \tau$.

To achieve $f_1(m) = f_2(m)$ we adjust $\omega_c$
so that the ions' separation $d \simeq 9\,\mu$m
is an integer number $p$ of standing wave periods, $\Delta k d =
2 \pi p$.
%This is done by driving the motion at $\omega = \omega_s$
%and finding the trap frequency where excitation of the stretch mode is minimised.
We used $p=22,\, \omega_c/2\pi = 500\,$kHz ($\eta=0.133$) in
one experiment, and $p=21,\, \omega_c/2\pi = 536.5\,$kHz ($\eta=0.128$)
in another.

A given experimental sequence consists of cooling, spin
preparation in $\ket{\downdown}$, then a spin echo sequence with
the force pulse $\cal W$ of duration $\tau$ in one or both of the
gaps, followed by an analysis pulse and then spin state
measurement; see inset to fig. 1b. The analysis pulse is the
rotation $R(\theta,\phi)$ in the tomography scheme.
A perfect implementation would
produce the maximally entangled state
\be
\ket{E(r)} \equiv (\ket{\upup} + e^{i r}\ket{\downdown})/\sqrt{2}
\ee
before the analysis pulse, with $r=2 \phi_1 - \pi/2$. To assess the
state $\rho$ obtained in practice we use the fidelity
$F \equiv {\rm max}_r \bra{E(r)}\rho\ket{E(r)}$. This compares
$\rho$ with the most closely matching member of the class $\ket{E(r)}$
\cite{footnote4}.

\begin{figure}[ht!]
\vspace{0cm}
\centerline{\resizebox{!}{0.5\textwidth}{\includegraphics{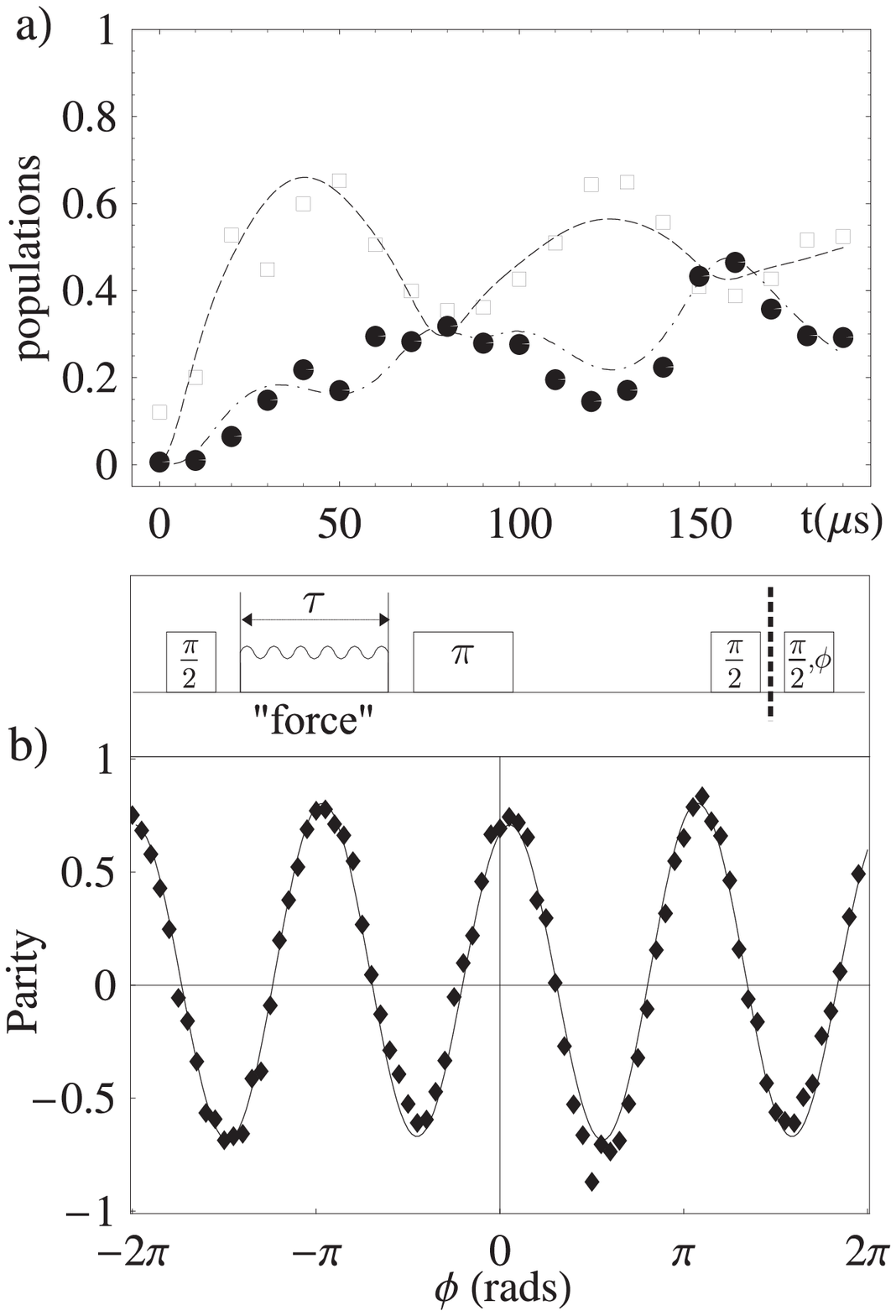}}}
\caption{a) Measured populations $P_R(\uparrow \uparrow)$, $P_R(\uparrow \downarrow)+P_R(\downarrow \uparrow)$ ($\bullet$, {\tiny$\Box$})
vs. $\tau$ after a spin-echo sequence with
${\cal W}(\tau)$ in the first gap, at $\delta/2\pi = 12.6$kHz. Each point is the average of 500 repeats of the experimental sequence.
The lines are fitted curves, using a model which assumes each qubit decoheres independently at rate $\Gamma$.
b) Parity signal vs. $\phi$ after a $\theta=0.46 \pi$ analysis
pulse, for $\tau=77\,\mu$s. Each point is the average of 1000 repeats of the experimental sequence. The inset shows the pulse sequence.}
\end{figure}

We use ``Schr\"odinger cat'' experiments with a single ion
\cite{96Monroe,ourcat} to analyse the forced motion and characterise
light shifts. We
then use the two-ion behaviour as a function of
$\tau$ with no analysis pulse to diagnose the setup.
We model the results by assuming each qubit decoheres independently at rate $\Gamma$.
For the case of a $\cal W$ pulse only in the first gap we then
expect
\be
P(\upup) &=& A - \mbox{\small $\frac{1}{2}$} e^{-\Gamma \tau-|\alpha(\tau)|^2/2}
\cos(\Psi(\tau)) \cos(\Delta_c \tau)  \nonumber\\
P(\updown) &+& P(\downup) \; =\; 1-2A \label{Pfit}
\ee
where $A = (1/4) + e^{-2\Gamma\tau}
[ \cos(2\Delta_c \tau) + e^{-2|\alpha(\tau)|^2} ]/8$ and $\alpha(\tau)$,
$\Psi(\tau)$ are the motional displacement and phase as described
in ref \cite{03Leibfried}.
Data is shown in fig 1a, fitted with floated parameters
$\Gamma,\delta,\Omega_f$ and $\Delta \phi$. We obtained $\Gamma=5.4\,$ms$^{-1}$, $\delta/2\pi =
13\,$kHz, $\Omega_f/2\pi = 23\,$kHz and $\Delta \phi = 1.6$ from the fit; these values were
consistent with our other information on $\delta$, $\Omega_c$ and $\Delta\phi$.
Note that $\Omega_f/\delta \simeq \sqrt{3}$ so in this experiment $\Psi=3\pi/2$.
When we use the model and fitted values to infer $\rho$
at $\tau=2\pi/\delta$, we obtain fidelity $F = 0.7(1)$.

We next measured a lower bound more directly, $F \ge 2 |C|$,
by using an analysis pulse
at $\theta=\pi/2$ to deduce the coherence $C =
\rho_{\upup,\downdown} = c_{11}-c_{22}+ i(c_{12}+c_{21})$. This
is obtained from the component at frequency 2 in the parity
signal $P_R(\updown)+P_R(\downup)$ vs. $\phi$ (data shown in fig. 1b.).
We observed $F \ge 0.74 (3)$.
This result is more precise than the previous one because it does not
rely on assumptions about the decoherence.

\begin{figure}[htb!]
\vspace{0 cm}
\centerline{\resizebox{!}{7.7cm}{\includegraphics{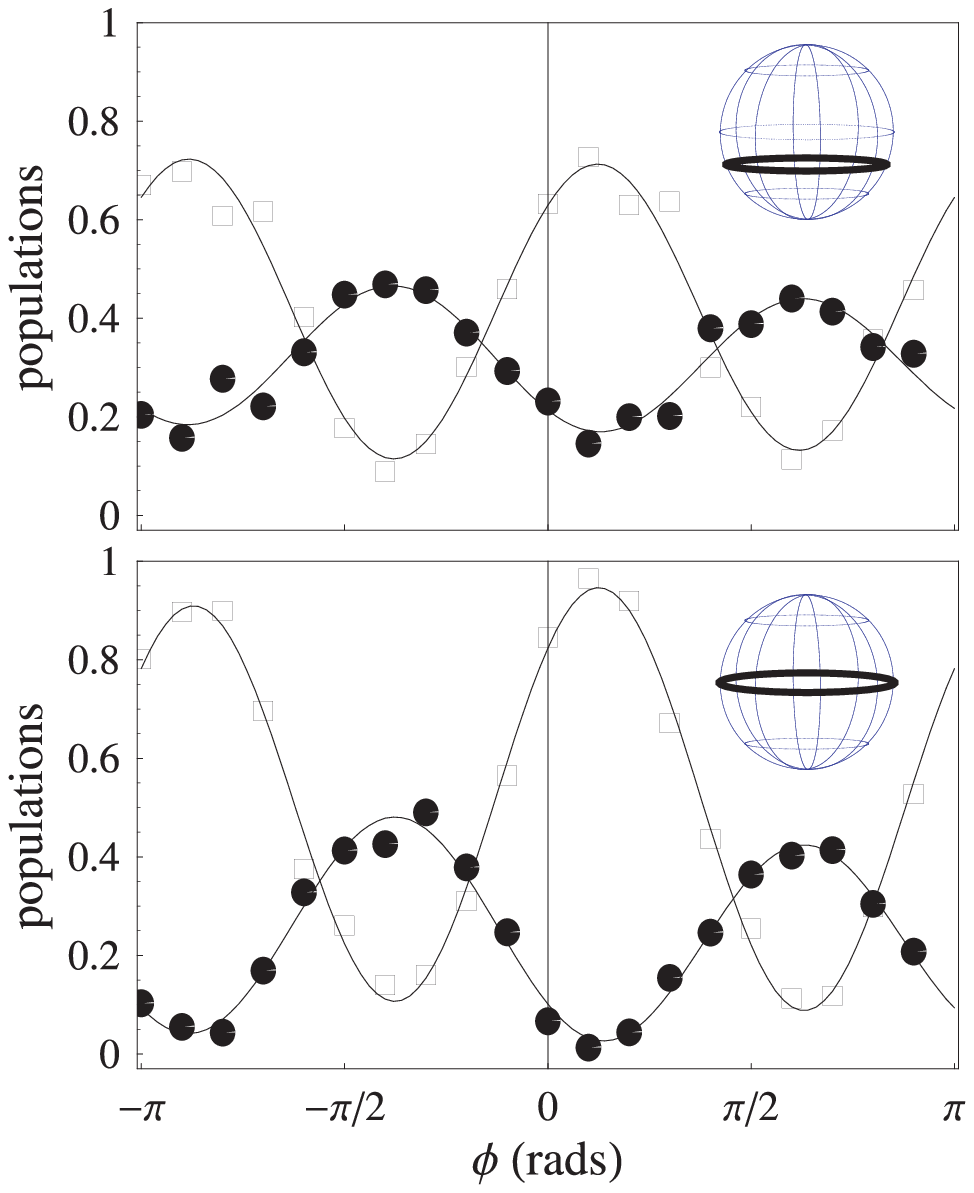}}}
\caption{Tomography signals for spin state after the spin echo
(${\cal W}(44\,\mu{\rm s})$ in both gaps,
$\delta/2\pi=22.7\,$kHz). The data points show $P_R(\upup),
P_R(\updown)+P_R(\downup)$ ($\bullet$, {\tiny$\Box$}) vs. $\phi$
for $\theta=0.66\pi$ (top) and $\theta=0.54\pi$ (bottom), as deduced from 500 repeats of the experimental sequence. The
lines are the fitted curves (\ref{fit}). $\theta$ values were
deduced from carrier flopping signals, taking into account a
$0.1\,\mu$s dead time in the AOM; $\phi$ is accurately known from
the r.f. signal generators. Inset: locus of $\theta,\phi$ values
on the Bloch sphere.} \label{fig:tomog}
\end{figure}

\begin{figure}[htb!]
\vspace*{0 cm}
\centerline{\resizebox{!}{5.0cm}{\includegraphics{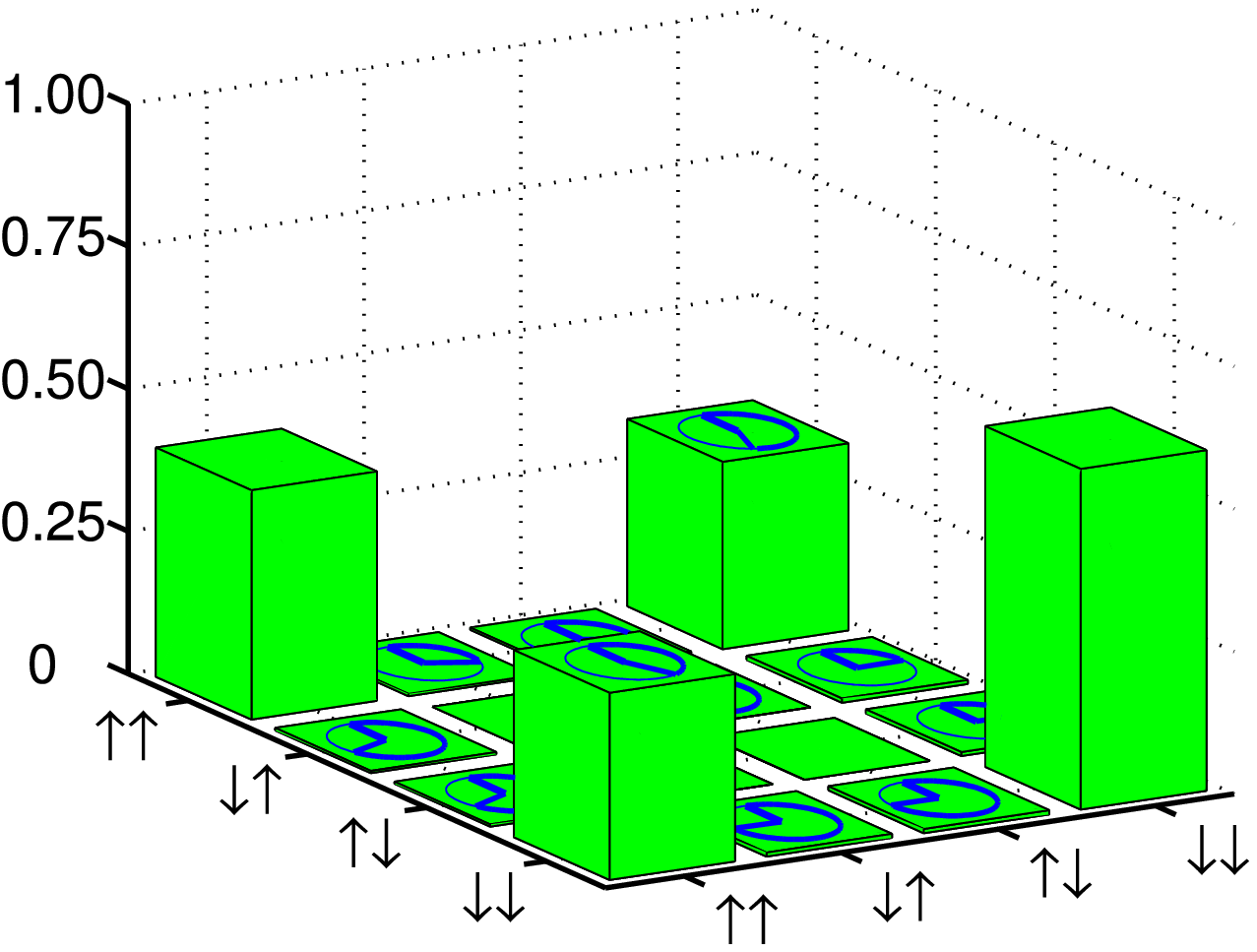}}}
\caption{Density matrix obtained from tomography data shown in fig. 2. The height of the bars indicates the absolute value of each density matrix element, and the clock faces indicate their phase. For this example the density matrix is closest to the state (3) with $r = 1.15 \pi$.}
\label{fig:densmat}
\end{figure}

Finally, we performed tomography in another experiment where the $\cal W$ pulse
was inserted in both gaps. This has the effect of cancelling the single-bit
rotations $Z_k(\phi_1)$. Both pulses had duration $\tau = 2\pi/\delta$ so the motion
completes 2 loops in phase space, and the gate requires $\Psi=\pi/4$ for
each loop. We set $\delta/2\pi = 22.7$ kHz and
observed the signal with pulses of double length in order to
adjust the laser intensity. We then used the carrier flopping rate to
infer $\Omega_f$, obtaining $2 \pi \times 16.3 (9)\,$kHz, giving the consistency
check $(\Omega_f/\delta)^2=0.52(5)$.

The data and fitted curves for the tomography are shown in fig. \ref{fig:tomog},
and the inferred (maximum likelihood) density matrix is shown in fig. \ref{fig:densmat}.
Owing to the absence of single ion addressing
in the rotations and the measurements, the tomography only gives
partial information on the internal elements of $\rho$ (i.e. away
from the corners in fig. \ref{fig:densmat}). However, we find that there is negligible
population in $\ket{\updown}$ and $\ket{\downup}$ and therefore for
the state in question, the tomography is complete. The inferred $\rho$
had fidelity $F=0.83 (2)$ and entanglement of formation \cite{BkNielsen} $0.54$.
The improvement is owing to better
precision in pulse areas and a reduction in the total
area by $\sqrt{2/3}$, which reduces decoherence by photon scattering.

We estimate that the primary source of infidelity is photon
scattering (12\%), with additional contributions from imbalance in
light intensity at the ions, pulse area imprecision, and motional
decoherence. Had we used a gate method which is more sensitive to
the prepared motional state, such as \cite{95Cirac}, the thermal
effect alone would contribute a further 6\% infidelity through
imprecise sideband pulses. We therefore obtained a useful
improvement by using a temperature-insensitive method.

This work was supported by the EPSRC, the Royal Society, the European Union, the National
Security Agency (NSA) and Advanced Research and Development Activity (ARDA) (P-43513-PH-QCO-02107-1)

\bibliographystyle{apsrev}

\begin{thebibliography}{25}
\expandafter\ifx\csname natexlab\endcsname\relax\def\natexlab#1{#1}\fi
\expandafter\ifx\csname bibnamefont\endcsname\relax
  \def\bibnamefont#1{#1}\fi
\expandafter\ifx\csname bibfnamefont\endcsname\relax
  \def\bibfnamefont#1{#1}\fi
\expandafter\ifx\csname citenamefont\endcsname\relax
  \def\citenamefont#1{#1}\fi
\expandafter\ifx\csname url\endcsname\relax
  \def\url#1{\texttt{#1}}\fi
\expandafter\ifx\csname urlprefix\endcsname\relax\def\urlprefix{URL }\fi
\providecommand{\bibinfo}[2]{#2}
\providecommand{\eprint}[2][]{\url{#2}}

\bibitem[{\citenamefont{Leibfried et~al.}(2003)\citenamefont{Leibfried,
  DeMarco, Meyer, Lucas, Barrett, Britton, Itano, Jelenkovic, Langer, Rosenband
  et~al.}}]{03Leibfried}
\bibinfo{author}{\bibfnamefont{D.}~\bibnamefont{Leibfried}},
  \bibinfo{author}{\bibfnamefont{B.}~\bibnamefont{DeMarco}},
  \bibinfo{author}{\bibfnamefont{V.}~\bibnamefont{Meyer}},
  \bibinfo{author}{\bibfnamefont{D.}~\bibnamefont{Lucas}},
  \bibinfo{author}{\bibfnamefont{M.}~\bibnamefont{Barrett}},
  \bibinfo{author}{\bibfnamefont{J.}~\bibnamefont{Britton}},
  \bibinfo{author}{\bibfnamefont{W.~M.} \bibnamefont{Itano}},
  \bibinfo{author}{\bibfnamefont{B.}~\bibnamefont{Jelenkovic}},
  \bibinfo{author}{\bibfnamefont{C.}~\bibnamefont{Langer}},
  \bibinfo{author}{\bibfnamefont{T.}~\bibnamefont{Rosenband}},
  \bibnamefont{et~al.}, \bibinfo{journal}{Nature}
  \textbf{\bibinfo{volume}{422}}, \bibinfo{pages}{412} (\bibinfo{year}{2003}).

\bibitem[{\citenamefont{Milburn et~al.}(2000)\citenamefont{Milburn, Schneider,
  and James}}]{00Milburn2}
\bibinfo{author}{\bibfnamefont{G.~J.} \bibnamefont{Milburn}},
  \bibinfo{author}{\bibfnamefont{S.}~\bibnamefont{Schneider}},
  \bibnamefont{and} \bibinfo{author}{\bibfnamefont{D.~F.} \bibnamefont{James}},
  \bibinfo{journal}{Fortschr. Physik} \textbf{\bibinfo{volume}{48}},
  \bibinfo{pages}{801} (\bibinfo{year}{2000}).

\bibitem[{\citenamefont{S{\o}rensen and M{\o}lmer}(2000)}]{00Sorensen1}
\bibinfo{author}{\bibfnamefont{A.}~\bibnamefont{S{\o}rensen}} \bibnamefont{and}
  \bibinfo{author}{\bibfnamefont{K.}~\bibnamefont{M{\o}lmer}},
  \bibinfo{journal}{Phys. Rev. A} \textbf{\bibinfo{volume}{62}},
  \bibinfo{pages}{022311} (\bibinfo{year}{2000}),
  \bibinfo{note}{quant-ph/0002024}.

\bibitem[{\citenamefont{Wang et~al.}(2001)\citenamefont{Wang, Sorensen, and
  Molmer}}]{01Wang}
\bibinfo{author}{\bibfnamefont{X.}~\bibnamefont{Wang}},
  \bibinfo{author}{\bibfnamefont{A.}~\bibnamefont{Sorensen}}, \bibnamefont{and}
  \bibinfo{author}{\bibfnamefont{K.}~\bibnamefont{Molmer}},
  \bibinfo{journal}{Phys. Rev. Lett.} \textbf{\bibinfo{volume}{86}},
  \bibinfo{pages}{3907} (\bibinfo{year}{2001}).

\bibitem[{\citenamefont{S{\o}rensen and M{\o}lmer}(1999)}]{99Sorensen}
\bibinfo{author}{\bibfnamefont{A.}~\bibnamefont{S{\o}rensen}} \bibnamefont{and}
  \bibinfo{author}{\bibfnamefont{K.}~\bibnamefont{M{\o}lmer}},
  \bibinfo{journal}{Phys. Rev. Lett.} \textbf{\bibinfo{volume}{82}},
  \bibinfo{pages}{1971} (\bibinfo{year}{1999}).

\bibitem[{\citenamefont{Turchette et~al.}(1998)\citenamefont{Turchette, Wood,
  King, Myatt, Leibfried, Itano, Monroe, and Wineland}}]{98Turchette}
\bibinfo{author}{\bibfnamefont{Q.~A.} \bibnamefont{Turchette}},
  \bibinfo{author}{\bibfnamefont{C.~S.} \bibnamefont{Wood}},
  \bibinfo{author}{\bibfnamefont{B.~E.} \bibnamefont{King}},
  \bibinfo{author}{\bibfnamefont{C.~J.} \bibnamefont{Myatt}},
  \bibinfo{author}{\bibfnamefont{D.}~\bibnamefont{Leibfried}},
  \bibinfo{author}{\bibfnamefont{W.~M.} \bibnamefont{Itano}},
  \bibinfo{author}{\bibfnamefont{C.}~\bibnamefont{Monroe}}, \bibnamefont{and}
  \bibinfo{author}{\bibfnamefont{D.~J.} \bibnamefont{Wineland}},
  \bibinfo{journal}{PRL} \textbf{\bibinfo{volume}{81}}, \bibinfo{pages}{3631}
  (\bibinfo{year}{1998}).

\bibitem[{\citenamefont{Haljan et~al.}(2005)\citenamefont{Haljan, Lee,
  Brickman, Acton, Deslauriers, and Monroe}}]{05Haljan}
\bibinfo{author}{\bibfnamefont{P.~C.} \bibnamefont{Haljan}},
  \bibinfo{author}{\bibfnamefont{P.~J.} \bibnamefont{Lee}},
  \bibinfo{author}{\bibfnamefont{K.-A.} \bibnamefont{Brickman}},
  \bibinfo{author}{\bibfnamefont{M.}~\bibnamefont{Acton}},
  \bibinfo{author}{\bibfnamefont{L.}~\bibnamefont{Deslauriers}},
  \bibnamefont{and} \bibinfo{author}{\bibfnamefont{C.}~\bibnamefont{Monroe}},
  \bibinfo{journal}{Phys. Rev. A} \textbf{\bibinfo{volume}{72}},
  \bibinfo{pages}{062316} (\bibinfo{year}{2005}).

\bibitem[{\citenamefont{Leibfried et~al.}(2005)\citenamefont{Leibfried, Knill,
  Seidelin, Britton, Blakestad, Chiaverini, Hume, Itano, Jost, Langer
  et~al.}}]{05Leibfried}
\bibinfo{author}{\bibfnamefont{D.}~\bibnamefont{Leibfried}},
  \bibinfo{author}{\bibfnamefont{E.}~\bibnamefont{Knill}},
  \bibinfo{author}{\bibfnamefont{S.}~\bibnamefont{Seidelin}},
  \bibinfo{author}{\bibfnamefont{J.}~\bibnamefont{Britton}},
  \bibinfo{author}{\bibfnamefont{R.~B.} \bibnamefont{Blakestad}},
  \bibinfo{author}{\bibfnamefont{J.}~\bibnamefont{Chiaverini}},
  \bibinfo{author}{\bibfnamefont{D.~B.} \bibnamefont{Hume}},
  \bibinfo{author}{\bibfnamefont{W.~M.} \bibnamefont{Itano}},
  \bibinfo{author}{\bibfnamefont{J.~D.} \bibnamefont{Jost}},
  \bibinfo{author}{\bibfnamefont{C.}~\bibnamefont{Langer}},
  \bibnamefont{et~al.}, \bibinfo{journal}{Nature}
  \textbf{\bibinfo{volume}{438}}, \bibinfo{pages}{639} (\bibinfo{year}{2005}).

\bibitem[{\citenamefont{Schmidt-Kaler et~al.}(2003)\citenamefont{Schmidt-Kaler,
  H{\"a}ffner, Riebe, Gulde, Lancaster, Deuschle, Becher, Eschner, and
  Blatt}}]{03SchmidtKaler}
\bibinfo{author}{\bibfnamefont{F.}~\bibnamefont{Schmidt-Kaler}},
  \bibinfo{author}{\bibfnamefont{H.}~\bibnamefont{H{\"a}ffner}},
  \bibinfo{author}{\bibfnamefont{M.}~\bibnamefont{Riebe}},
  \bibinfo{author}{\bibfnamefont{S.}~\bibnamefont{Gulde}},
  \bibinfo{author}{\bibfnamefont{G.~P.~T.} \bibnamefont{Lancaster}},
  \bibinfo{author}{\bibfnamefont{T.}~\bibnamefont{Deuschle}},
  \bibinfo{author}{\bibfnamefont{C.}~\bibnamefont{Becher}},
  \bibinfo{author}{\bibfnamefont{C.~F. R.~J.} \bibnamefont{Eschner}},
  \bibnamefont{and} \bibinfo{author}{\bibfnamefont{R.}~\bibnamefont{Blatt}},
  \bibinfo{journal}{Nature} \textbf{\bibinfo{volume}{422}}, \bibinfo{pages}{408
  } (\bibinfo{year}{2003}).

\bibitem[{\citenamefont{H\"{a}ffner et~al.}(2005)\citenamefont{H\"{a}ffner,
  H{\"a}nsel, Roos, Benhelm, al~kar, Chwalla, K\"{o}rber, Rapol, Riebe, Schmidt
  et~al.}}]{05Haffner}
\bibinfo{author}{\bibfnamefont{H.}~\bibnamefont{H\"{a}ffner}},
  \bibinfo{author}{\bibfnamefont{W.}~\bibnamefont{H{\"a}nsel}},
  \bibinfo{author}{\bibfnamefont{C.~F.} \bibnamefont{Roos}},
  \bibinfo{author}{\bibfnamefont{J.}~\bibnamefont{Benhelm}},
  \bibinfo{author}{\bibfnamefont{D.~C.} \bibnamefont{al~kar}},
  \bibinfo{author}{\bibfnamefont{M.}~\bibnamefont{Chwalla}},
  \bibinfo{author}{\bibfnamefont{T.}~\bibnamefont{K\"{o}rber}},
  \bibinfo{author}{\bibfnamefont{U.~D.} \bibnamefont{Rapol}},
  \bibinfo{author}{\bibfnamefont{M.}~\bibnamefont{Riebe}},
  \bibinfo{author}{\bibfnamefont{P.~O.} \bibnamefont{Schmidt}},
  \bibnamefont{et~al.}, \bibinfo{journal}{Nature}
  \textbf{\bibinfo{volume}{438}}, \bibinfo{pages}{643} (\bibinfo{year}{2005}).
  
\bibitem{footnote1}
Ion spin qubits
are entangled in H. H\"{a}ffner {\em et al.}, Appl. Phys. B. {\bf 81},151 (2005),
by transfer from an $S_{1/2}/D_{5/2}$ qubit.

\bibitem[{\citenamefont{Lo et~al.}(1998)\citenamefont{Lo, Popescu, and
  Spiller}}]{BkLo}
\bibinfo{editor}{\bibfnamefont{H.-K.} \bibnamefont{Lo}},
  \bibinfo{editor}{\bibfnamefont{S.}~\bibnamefont{Popescu}}, \bibnamefont{and}
  \bibinfo{editor}{\bibfnamefont{T.}~\bibnamefont{Spiller}}, eds.,
  \emph{\bibinfo{title}{Introduction to quantum computation and information}}
  (\bibinfo{publisher}{World Scientific}, \bibinfo{address}{Singapore},
  \bibinfo{year}{1998}).

\bibitem[{\citenamefont{Nielsen and Chuang}(2000)}]{BkNielsen}
\bibinfo{author}{\bibfnamefont{M.~A.} \bibnamefont{Nielsen}} \bibnamefont{and}
  \bibinfo{author}{\bibfnamefont{I.~L.} \bibnamefont{Chuang}},
  \emph{\bibinfo{title}{Quantum Computation and Quantum Information}}
  (\bibinfo{publisher}{Cambridge University Press},
  \bibinfo{address}{Cambridge}, \bibinfo{year}{2000}).

\bibitem[{\citenamefont{Lucas et~al.}(2004)\citenamefont{Lucas, Ramos, Home,
  J.McDonnell, Nakayama, Stacey, Webster, Stacey, and Steane}}]{04Lucas}
\bibinfo{author}{\bibfnamefont{D.~M.} \bibnamefont{Lucas}},
  \bibinfo{author}{\bibfnamefont{A.}~\bibnamefont{Ramos}},
  \bibinfo{author}{\bibfnamefont{J.~P.} \bibnamefont{Home}},
  \bibinfo{author}{\bibfnamefont{M.}~\bibnamefont{J.McDonnell}},
  \bibinfo{author}{\bibfnamefont{S.}~\bibnamefont{Nakayama}},
  \bibinfo{author}{\bibfnamefont{J.-P.} \bibnamefont{Stacey}},
  \bibinfo{author}{\bibfnamefont{S.~C.} \bibnamefont{Webster}},
  \bibinfo{author}{\bibfnamefont{D.~N.} \bibnamefont{Stacey}},
  \bibnamefont{and} \bibinfo{author}{\bibfnamefont{A.~M.}
  \bibnamefont{Steane}}, \bibinfo{journal}{Phys. Rev. A}
  \textbf{\bibinfo{volume}{69}}, \bibinfo{pages}{012711}
  (\bibinfo{year}{2004}).

\bibitem[{\citenamefont{Turchette et~al.}(2000)\citenamefont{Turchette,
  Kielpinski, King, Leibfried, Meekhof, Myatt, Rowe, Sackett, Wood, Itano
  et~al.}}]{00Turchette}
\bibinfo{author}{\bibfnamefont{Q.~A.} \bibnamefont{Turchette}},
  \bibinfo{author}{\bibfnamefont{D.}~\bibnamefont{Kielpinski}},
  \bibinfo{author}{\bibfnamefont{B.~E.} \bibnamefont{King}},
  \bibinfo{author}{\bibfnamefont{D.}~\bibnamefont{Leibfried}},
  \bibinfo{author}{\bibfnamefont{D.~M.} \bibnamefont{Meekhof}},
  \bibinfo{author}{\bibfnamefont{C.~J.} \bibnamefont{Myatt}},
  \bibinfo{author}{\bibfnamefont{M.~A.} \bibnamefont{Rowe}},
  \bibinfo{author}{\bibfnamefont{C.~A.} \bibnamefont{Sackett}},
  \bibinfo{author}{\bibfnamefont{C.~S.} \bibnamefont{Wood}},
  \bibinfo{author}{\bibfnamefont{W.~M.} \bibnamefont{Itano}},
  \bibnamefont{et~al.}, \bibinfo{journal}{Phys. Rev. A}
  \textbf{\bibinfo{volume}{61}}, \bibinfo{pages}{063418}
  (\bibinfo{year}{2000}).

\bibitem[{\citenamefont{Deslauriers et~al.}(2004)\citenamefont{Deslauriers,
  Haljan, Lee, Brickman, Blinov, Madsen, and Monroe}}]{04Deslauriers}
\bibinfo{author}{\bibfnamefont{L.}~\bibnamefont{Deslauriers}},
  \bibinfo{author}{\bibfnamefont{P.~C.} \bibnamefont{Haljan}},
  \bibinfo{author}{\bibfnamefont{P.~J.} \bibnamefont{Lee}},
  \bibinfo{author}{\bibfnamefont{K.-A.} \bibnamefont{Brickman}},
  \bibinfo{author}{\bibfnamefont{B.~B.} \bibnamefont{Blinov}},
  \bibinfo{author}{\bibfnamefont{M.~J.} \bibnamefont{Madsen}},
  \bibnamefont{and} \bibinfo{author}{\bibfnamefont{C.}~\bibnamefont{Monroe}},
  \bibinfo{journal}{Phys. Rev. A}  (\bibinfo{year}{2004}),
  \bibinfo{note}{quant-ph/0404142}.

\bibitem[{\citenamefont{Wineland et~al.}(1998)\citenamefont{Wineland, Monroe,
  Itano, Leibfried, King, and Meekhof}}]{98Wineland2}
\bibinfo{author}{\bibfnamefont{D.~J.} \bibnamefont{Wineland}},
  \bibinfo{author}{\bibfnamefont{C.}~\bibnamefont{Monroe}},
  \bibinfo{author}{\bibfnamefont{W.~M.} \bibnamefont{Itano}},
  \bibinfo{author}{\bibfnamefont{D.}~\bibnamefont{Leibfried}},
  \bibinfo{author}{\bibfnamefont{B.~E.} \bibnamefont{King}}, \bibnamefont{and}
  \bibinfo{author}{\bibfnamefont{D.~M.} \bibnamefont{Meekhof}},
  \bibinfo{journal}{J. Res. Natl. Inst. Stand. Technol.}
  \textbf{\bibinfo{volume}{103}}, \bibinfo{pages}{259} (\bibinfo{year}{1998}).

\bibitem[{\citenamefont{Kielpinski et~al.}(2002)\citenamefont{Kielpinski,
  C.Monroe, and Wineland}}]{02Kielpinski}
\bibinfo{author}{\bibfnamefont{D.}~\bibnamefont{Kielpinski}},
  \bibinfo{author}{\bibnamefont{C.Monroe}}, \bibnamefont{and}
  \bibinfo{author}{\bibfnamefont{D.}~\bibnamefont{Wineland}},
  \bibinfo{journal}{Nature} \textbf{\bibinfo{volume}{417}},
  \bibinfo{pages}{709} (\bibinfo{year}{2002}).

\bibitem[{\citenamefont{Steane}(2004)}]{04Steane2}
\bibinfo{author}{\bibfnamefont{A.}~\bibnamefont{Steane}}
  (\bibinfo{year}{2004}), \bibinfo{note}{quant-ph/0412165}.

\bibitem[{\citenamefont{et. al.}(2006)}]{ourcat}
\bibinfo{author}{\bibfnamefont{M.~J.~M.} \bibnamefont{et. al.}},
  \bibinfo{journal}{(in preparation)}  (\bibinfo{year}{2006}).

\bibitem{footnote2}
The motion leads
to a two-qubit effect, $\Psi \ne 0$, only if $\omega_s
\ne \omega_c$. Since this difference is owing to the ions'
Coulomb repulsion, the 2-qubit gate
relies, as expected, on a two-body interaction.  

\bibitem[{\citenamefont{Paris et~al.}(2004)\citenamefont{Paris, Matteo,
  Rehacek, and Jaroslav}}]{BkParis}
\bibinfo{editor}{\bibnamefont{Paris}}, \bibinfo{editor}{\bibnamefont{Matteo}},
  \bibinfo{editor}{\bibnamefont{Rehacek}}, \bibnamefont{and}
  \bibinfo{editor}{\bibnamefont{Jaroslav}}, eds., \emph{\bibinfo{title}{Quantum
  state estimation}} (\bibinfo{publisher}{Springer}, \bibinfo{address}{Berlin},
  \bibinfo{year}{2004}).

\bibitem[{\citenamefont{James et~al.}(2001)\citenamefont{James, Kwiat, Munro,
  and White}}]{01James}
\bibinfo{author}{\bibfnamefont{D.~F.~V.} \bibnamefont{James}},
  \bibinfo{author}{\bibfnamefont{P.~G.} \bibnamefont{Kwiat}},
  \bibinfo{author}{\bibfnamefont{W.~J.} \bibnamefont{Munro}}, \bibnamefont{and}
  \bibinfo{author}{\bibfnamefont{A.~G.} \bibnamefont{White}},
  \bibinfo{journal}{Phys. Rev. A} \textbf{\bibinfo{volume}{64}}
  (\bibinfo{year}{2001}).

\bibitem[{\citenamefont{McDonnell et~al.}(2004)\citenamefont{McDonnell, Stacey,
  Webster, Home, Ramos, Lucas, Stacey, and Steane}}]{04McDonnell1}
\bibinfo{author}{\bibfnamefont{M.}~\bibnamefont{McDonnell}},
  \bibinfo{author}{\bibfnamefont{J.-P.} \bibnamefont{Stacey}},
  \bibinfo{author}{\bibfnamefont{S.~C.} \bibnamefont{Webster}},
  \bibinfo{author}{\bibfnamefont{J.~P.} \bibnamefont{Home}},
  \bibinfo{author}{\bibfnamefont{A.}~\bibnamefont{Ramos}},
  \bibinfo{author}{\bibfnamefont{D.~M.} \bibnamefont{Lucas}},
  \bibinfo{author}{\bibfnamefont{D.~N.} \bibnamefont{Stacey}},
  \bibnamefont{and} \bibinfo{author}{\bibfnamefont{A.~M.}
  \bibnamefont{Steane}}, \bibinfo{journal}{Phys. Rev. Lett.}
  \textbf{\bibinfo{volume}{93}}, \bibinfo{pages}{153601}
  (\bibinfo{year}{2004}).

\bibitem[{\citenamefont{King et~al.}(1998)\citenamefont{King, Wood, Myatt,
  Turchette, Leibfried, Itano, Monroe, and Wineland}}]{98King}
\bibinfo{author}{\bibfnamefont{B.~E.} \bibnamefont{King}},
  \bibinfo{author}{\bibfnamefont{C.~S.} \bibnamefont{Wood}},
  \bibinfo{author}{\bibfnamefont{C.~J.} \bibnamefont{Myatt}},
  \bibinfo{author}{\bibfnamefont{Q.~A.} \bibnamefont{Turchette}},
  \bibinfo{author}{\bibfnamefont{D.}~\bibnamefont{Leibfried}},
  \bibinfo{author}{\bibfnamefont{W.~M.} \bibnamefont{Itano}},
  \bibinfo{author}{\bibfnamefont{C.}~\bibnamefont{Monroe}}, \bibnamefont{and}
  \bibinfo{author}{\bibfnamefont{D.~J.} \bibnamefont{Wineland}},
  \bibinfo{journal}{PRL} \textbf{\bibinfo{volume}{81}}, \bibinfo{pages}{1525}
  (\bibinfo{year}{1998}).
  
\bibitem{footnote3}
Beam $A$ drives the carrier by a Raman process off-resonant
by $\omega_0$; this creates a light shift whose cancellation results in
$\epsilon_- / \epsilon_+ \simeq 1.02$ for this beam.

\bibitem{footnote4}
This choice is reasonable because it is equivalent to
using the data to extract the value of $r$.

\bibitem[{\citenamefont{Monroe et~al.}(1996)\citenamefont{Monroe, Meekhof,
  King, and Wineland}}]{96Monroe}
\bibinfo{author}{\bibfnamefont{C.}~\bibnamefont{Monroe}},
  \bibinfo{author}{\bibfnamefont{D.~M.} \bibnamefont{Meekhof}},
  \bibinfo{author}{\bibfnamefont{B.~E.} \bibnamefont{King}}, \bibnamefont{and}
  \bibinfo{author}{\bibfnamefont{D.~J.} \bibnamefont{Wineland}},
  \bibinfo{journal}{Science} \textbf{\bibinfo{volume}{272}},
  \bibinfo{pages}{1131} (\bibinfo{year}{1996}).

\bibitem[{\citenamefont{Cirac and Zoller}(1995)}]{95Cirac}
\bibinfo{author}{\bibfnamefont{J.~I.} \bibnamefont{Cirac}} \bibnamefont{and}
  \bibinfo{author}{\bibfnamefont{P.}~\bibnamefont{Zoller}},
  \bibinfo{journal}{Phys. Rev. Lett.} \textbf{\bibinfo{volume}{74}},
  \bibinfo{pages}{4091} (\bibinfo{year}{1995}).

\end{thebibliography}

\end{document}